\documentstyle[12pt,aaspp,psfig]{article}

\def\Msun{\hbox{$\rm\thinspace M_{\odot}$}}

\received{ }
\accepted{1998 Jan 28}
\cpright{ }{1998}
\slugcomment{To appear in Astrophysical Journal, 1998 Vol. 499.}

\begin{document}

\title{Gamma-Ray Spectral States of Galactic Black Hole Candidates}

\author{ J.E. Grove\altaffilmark{1}, W.N. Johnson, R.A. Kroeger, 
K. McNaron-Brown\altaffilmark{2}, J.G. Skibo}
\affil{E.O. Hulburt Center for Space Research, Code 7650, Naval
Research Lab., Washington DC 20375}
\author{ B.F. Phlips }
\affil{Universities Space Research Association, Washington DC 20024}
\altaffiltext{1}{E-mail:  grove@osse.nrl.navy.mil}
\altaffiltext{2}{CSI, George Mason University, Fairfax, VA 22030}

\begin{abstract}
OSSE has observed seven transient black hole candidates: GRO~J0422+32,
GX339--4, GRS~1716--249, GRS~1009--45, 4U~1543--47, GRO~J1655--40, and
GRS~1915+105.  Two gamma-ray spectral states are evident and, based on
a limited number of contemporaneous X-ray and gamma-ray observations,
these states appear to be correlated with X-ray states.  The former
three objects show hard spectra below 100 keV (photon number indices
$\Gamma < 2$) that are exponentially cut off with folding energy
$\sim$ 100 keV, a spectral form that is consistent with thermal
Comptonization.  This ``breaking gamma-ray state'' is the high-energy
extension of the X-ray low, hard state.  In this state, the
majority of the luminosity is above the X-ray band, carried by photons
of energy $\sim$100 keV.  The latter four
objects exhibit a ``power-law gamma-ray state'' with a relatively soft
spectral index ($\Gamma \sim 2.5-3$) and no evidence for a
spectral break.  For GRO~J1655--40, the
lower limit on the break energy is 690 keV.  GRS~1716--249 exhibits 
both spectral states, with the power-law state having significantly
lower gamma-ray luminosity.  The power-law gamma-ray state is associated
with the presence of a strong ultrasoft X-ray excess (kT $\sim$ 1 keV), the
signature of the X-ray high, soft (or perhaps very high) state.  The 
physical process responsible for the unbroken power law is not well 
understood, although the spectra are consistent with bulk-motion
Comptonization in the convergent accretion flow.
\end{abstract}


\section {Introduction}

The question of whether neutron stars and black holes can be distinguished
by their X-ray and gamma-ray spectra is important, and bears on
the classification of newly discovered transients.  Before such distinctions
can be made reliably, the full range of spectral forms must be observed and
categorized.  Extensive knowledge of the X-ray emission of these objects
has accumulated in the literature, but the nature of the gamma-ray emission
is only now coming to light.

The historical record of X-ray (i.e. $<$30 keV) observations 
of galactic black hole candidates (BHCs)
reveals at least four spectral states, listed here in order of decreasing
X-ray luminosity (see, e.g., Tanaka 1989, Grebenev et al. 1993, and van der
Klis 1994, 1995).  In the ``{\em X-ray very high}'' state, the soft (i.e. 1--10
keV) X-ray flux is quite high, with an ``ultrasoft'' thermal or multi-color
blackbody spectrum of characteristic temperature kT $\sim$ 1 keV; a
power-law tail is present with a photon
number index $\Gamma \sim $ 2--3.  The typical 
X-ray luminosity is approximately at the Eddington 
limit.  Rapid intensity variations are observed and are associated
with the power-law tail.  The amplitude of this rapid variability is
less than that in the X-ray low, hard state but significantly stronger
than in the X-ray high, soft state (see below).  In the 
``{\em X-ray high, soft}'' state, the spectrum again shows both an ultrasoft
component with kT $\sim$ 1 keV and a weak power-law tail that dominates
above $\sim$10 keV.  This state is
distinguished from the X-ray very high state by its lower luminosity
(typically by a factor of $\sim$3--30) and weakness or absence of rapid
intensity variations.  The ``{\em X-ray low, hard}'' state exhibits a single
power-law spectrum with photon number index $\Gamma \sim $ 1.5--2, with
a typical X-ray luminosity of $<$1\% of Eddington.  (However, we show
below that in this state
the majority of the luminosity is carried by photons above
30 keV).  Strong
rapid intensity variability is observed, with rms variations
of order a few tens of percent of the total emission.  The ``{\em X-ray
off}'' or ``{\em quiescent}'' state exhibits very low level emission with
uncertain spectral shape at a luminosity $L_X < 10^{-4}$ of Eddington.

Here we summarize the low-energy gamma-ray spectra of seven transient 
galactic BHCs, all low-mass X-ray binaries (LMXBs), 
observed with OSSE on the Compton Gamma Ray Observatory
(in order of increasing right ascension:  GRO~J0422+32, 
GRS~1009--45, 4U~1543--47, GRO~J1655--40, GX339--4, 
GRS~1716--249, and GRS~1915+105) and demonstrate the existence of two spectral
states in low-energy gamma rays.  We compare these spectra with those
of the archetypal galactic BHC, the high-mass
X-ray binary Cyg X-1.  
Introductory remarks covering the historical
record of each of these objects are given along with the details
of the OSSE observations in the subsections below.

\section{Observations}

The OSSE instrument on the Compton GRO 
consists of four nearly identical large-area
NaI(Tl)--CsI(Na) phoswich detector systems (Johnson et al. 1993).  It can
be configured to cover
the energy range from $\simeq$40 keV to 10 MeV with good spectral resolution.
Simultaneously with the spectroscopy data, count-rate samples in a number of
energy bands can be collected on time scales of 4--32 ms to study rapid
variability.  Hard X-ray transients, including those later classified as 
BHCs, have been high-priority targets for OSSE throughout
the Compton GRO mission.  They have generally been observed as targets of
opportunity in response to
detection of a significant flaring event by the BATSE instrument on the
Compton GRO.  These target-of-opportunity pointings have lasted from less than
24 hours for GRS~1009--45 to a few weeks for GRO~J0422+32, depending on the 
strength and duration of the outburst,
as well as the flexibility of the Compton observing program.
Table \ref{dates} summarizes the dates on which OSSE observed the seven
gamma-ray transients discussed here.

Hard X-ray lightcurves (20--100 keV) from BATSE for each
transient in this study are shown in Fig. 
\ref{lightcurves}.  Each data point corresponds to a single day.
Note that one of two different flux scales is used in each panel,
depending on the intensity of the emission.  The total Crab nebular
and pulsar emission in this band is $\simeq$0.32 ph cm$^{-2}$ s$^{-1}$;
thus the outbursts studied here peak between about $1/3$ and 1.5 Crab
flux units.  OSSE observing periods 
are indicated by shaded vertical intervals
in each panel.  Note also that the lightcurves end on
MJD 49800.0 (1995 Mar 24.0), the last date for which BATSE flux measurements
are publicly available as of this writing.  For some sources, this is
before the last OSSE observation included in this study, and indeed before
the first OSSE observation.  The
lightcurves, nevertheless, serve to place the OSSE measurements in
the context of the recent historical behavior of each source.

\subsection{ GRO~J0422+32 }

GRO~J0422+32 (XN Per 1992) was discovered in outburst by BATSE in
data from 1992 August 5.  The lightcurve of the outburst showed a fast
rise and approximately exponential decay with $\tau \simeq 40$ days,
and a secondary maximum beginning $\simeq$125 days after the onset
of the initial outburst (Harmon et al. 1994).
It was observed in X-rays at various times in the outburst
by ASCA (Tanaka 1993a), ROSAT (Pietsch et al. 1993),
and Mir/TTM (Sunyaev et al. 1993).  The X-ray spectrum in all
cases was consistent with a simple power law,
with no evidence for an ultrasoft component.  While the mass function
of $1.2 \pm 0.04 \Msun$ determined by Filippenko, Matheson, \& Ho (1995)
is low enough that the compact object might indeed be a neutron star,
the H$\alpha$ radial velocity curve and the M stellar type of the
secondary imply a mass of 3.6$\Msun$ for the primary.  The photometric
measurements of Callanan et al. (1996) support this mass estimate
and give a distance of $\sim$2 kpc.  It was a transient radio
source, exhibiting a radio lightcurve consistent with a synchroton bubble
ejection event (Shrader et al. 1994).

OSSE observed GRO~J0422+32 for 33 days spanning the interval from the peak
of the outburst through the decline to approximately half maximum intensity
at 100 keV.  There was strong, rapid intensity variability above 50 keV
(rms $\simeq$ 30--40\% of the total emission)
and a substantial peaked noise component in the power spectrum
(Grove et al. 1994).  The photon
spectrum was described well by the simple, exponentially
truncated power-law form in Eqn. 1.  While this spectral shape was valid
for each observation day, there was substantial evolution of the model
parameters throughout the outburst.  As the intensity at 100 keV declined,
the exponential folding energy increased by about 20\%, with a nearly
linear anticorrelation between these parameters. 
The TTM and HEXE instruments on Mir-Kvant observed 1992 Aug 29 -- Sep 2
and reported a hard spectrum extending well above 100~keV
(Sunyaev et al. 1993; Maisack et al. 1994).  To create a consistent 
broadband spectrum, we therefore selected OSSE data only from
a four-day period surrounding this interval and fit the combined spectrum
(Table \ref{fits}).

\subsection{ GRS~1009--45}

The transient GRS 1009--45 (XN Vel 1993) was discovered by GRANAT/WATCH
on 1993 September 12 (Lapshov, Sazonov, \& Sunyaev 1993).  The lightcurve
above 20 keV for the discovery outburst
had a fast rise and exponential decay with $\tau \simeq 5$
days (Harmon et al. 1994).  The secondary outbursts, peaking $\sim$35 days
and $\sim$85 days after the initial outburst, had substantially longer
rise and fall times (see Fig. \ref{lightcurves}).
The X-ray spectrum was ultrasoft (Kaniovsky, Borozdin, \& Sunyaev 1993; 
Tanaka 1993b; Moss 1997; Ebisawa 1997).  A blue optical counterpart
has been identified (Della Valle \& Benetti 1993), of type late-G or 
early-K in a $6.86 \pm 0.12$ h binary orbit (Shahbaz et al. 1996).
A mass function has yet to be determined.  The distance is estimated
to be between 1.5 and 4.5 kpc.

OSSE observed GRS 1009--45 for a single 17-hour period
$\simeq$9 days after the onset of the outburst, and $\simeq$6 days 
after the peak, by
which time the 20--100 keV flux had decayed to $\sim$1/3 its maximum
(Harmon et al. 1994).  Despite the brevity of the pointing, OSSE 
clearly detected emission above 300 keV, with a power law spectrum
and no evidence for a break.  ASCA data from 1993 November 11 are shown
in Fig. \ref{spectra} together with the OSSE spectrum.  The ASCA data,
which have been taken from Moss (1997),
have been scaled upward by a factor of $\sim$30, corresponding to the
decline in the BATSE hard X-ray flux between the OSSE and ASCA observations.

\subsection{4U~1543--47}

BATSE detected 4U~1543--47 on 1992 April 18 in its third known outburst
since the discovery in 1971 (Matilsky et al. 1972).  The X-ray spectrum
in the discovery outburst was ultrasoft.  Lightcurves show
a fast rise and exponential decay, and multiple secondary maxima have
been observed.  The e-folding time of the decline of the X-ray emission
from the discovery outburst was $\tau \simeq$ 85 days, while BATSE measured
a much shorter $\tau \simeq$ 2.5 days above 20 keV in 1992.  A similar 
difference
in soft and hard X-ray decay times was observed from GS~1124--68
(Gilfanov et al. 1991).  ROSAT observed 4U~1543--47 in 1992 August and
September (Greiner et al. 1993), when the source was undetectable
by BATSE.  The X-ray spectrum was again ultrasoft.  There are no reports
in the literature of rapid intensity variations.  The optical counterpart
is an A-type dwarf at $\sim$4 kpc in an unknown binary orbit (Chevalier 1989).

OSSE observed 4U~1543--47 for a nine-day period beginning about 10 days
after its detection by BATSE.  The outburst peaked on
April 19, and had decayed to undetectability by BATSE by the time the
OSSE observation began.  The OSSE data indicate that the source
underwent a secondary outburst during this observation, at a flux level
below BATSE's daily sensitivity.  Statistically significant emission was 
detected by OSSE on the first six observing days, 1992 April 28 -- May 3.
The spectrum was a simple power law,
with emission detected to at least 200 keV.  While the gamma-ray luminosity
varied by a factor of $\sim$3, there was no statistically 
significant variation in the power-law index from day to day.

\subsection{GRO~J1655--40}

The transient X-ray source GRO~J1655--40 (XN Sco 1994) was discovered
by BATSE in data from 1994 July 27.  Since the discovery outburst, the
emission has been episodic, showing periods of weeks of strong, relatively
constant emission, then returning to quiescence (Harmon et al. 1995).
The mass function is measured to be 3.35$\pm$0.14 $\Msun$, and
optical eclipses are observed with a 2.6-day period, suggesting that
the system is viewed nearly edge-on (Bailyn et al. 1995a,b).  Orosz et
al. (1996) estimate the mass of the compact object to be 7$\Msun$.  It is a
strong, transient radio source with
superluminal jets (Tingay et al. 1995, Hjellming \& Rupen 1995) at a
distance of 3.2 kpc (McKay et al. 1994, Hjellming \& Rupen 1995,
Bailyn et al. 1995b).  In at least three cases, the X-ray outbursts
were followed within days by ejection events in the radio jets.

This source has been observed by OSSE on six occasions.
Results from the first five were presented by Kroeger et al. (1996).
Four observations provided strong detections of a
soft power-law spectrum, while in the remaining two observations the
source was undetectable by OSSE.  The dates of
the former are given in Table \ref{dates}.  A simple power-law
model gives a good fit to each observation, as well as to the sum
of all observations with high gamma-ray luminosity, $L_\gamma$.  Table 
\ref{fits} reports 
the photon index for the sum of the high $L_\gamma$
observations.  The envelope of the scatter of 
spectral index for the individual observations is 0.4.  Investigation of
the spectral index on $\sim$90-minute timescales reveals a weak
dependence of index with $L_\gamma$:  higher luminosity intervals tend
to have more negative spectral index (i.e. a steeper spectrum).  ASCA observed
GRO~J1655--40 on 1995 Aug 15-16, during an outburst that unfortunately was
not observed by OSSE.  The BATSE instrument, with its all-sky
capability, monitored this outburst in its entirety.  Combined ASCA and
BATSE spectra show an ultrasoft excess and a soft power-law tail
($\Gamma \simeq 2.4$) extending beyond 100 keV (Zhang et al. 1997).  The
ASCA data plotted in Fig. \ref{spectra} have been scaled by the ratio of
the average OSSE flux at 40 keV to the reported BATSE flux at 40 keV for
1995 Aug 15-16 (BATSE data are not shown in the figure).  The ultrasoft
excess remains apparent.  
During the most recent OSSE observation of GRO~J1655--40 (1996 Aug-Sep),
the source was also observed by the Rossi XTE, so high-quality, 
simultaneous, broadband spectra should soon be available.

\subsection{GX339--4}

The highly variable
black hole candidate GX339--4 (1H 1659-48.7) exhibits 
all four of the X-ray states (Miyamoto et al.
1991 and references therein).  Its optical counterpart (Doxsey et 
al. 1979) lies at a distance generally assumed to be $\sim$4 kpc.  It
is a variable radio source (Sood \& Campbell-Wilson 1994), and recent
evidence suggests it may have a jet (Fender et al. 1997).  In 
the X-ray low, hard state, rapid fluctuations
with rms variations as high as 30\% have been observed, as well as
QPOs near $\sim$0.8 Hz (Grebenev et al. 1991), $\sim$0.1 Hz and $\sim$0.05 Hz
(Motch et al. 1983).  Observations by Ginga in 1991 September
established that the source was in the X-ray low, hard state.  Contemporaneous
OSSE observations revealed that the low-energy gamma-ray emission was
strong and that the spectrum was exponentially cut off
(see Fig. \ref{spectra} and Table \ref{fits}), from which Grabelsky et al.
(1995) argued that the X-ray low, hard state corresponds to gamma-ray
outburst in this object.  Observations performed with GRANAT 
(Grebenev et al. 1993) indeed
show that GX339--4 exhibits distinct hard X-ray states correlated with the
X-ray behavior.

OSSE has observed GX339--4 four times.  The first pointing, in 1991
September, was made as
a target of opportunity in response to an outburst detected by BATSE.
The second observation was performed in 1991 November as part of the 
scheduled viewing plan, when the source happened to have very low
gamma-ray luminosity (Fig. \ref{lightcurves}).  Because of the
potential for confusion with diffuse continuum emission from the
galactic plane, we have not included these data in our analysis.
The final two observations contain proprietary data and have not
been included in our analysis or in Table \ref{dates}.

\subsection{GRS~1716--249}

The transient GRS 1716--249 (GRO~J1719--24, XN Oph 1993) was discovered
by GRANAT/SIGMA on 1993 September 25 (Ballet et al. 1993).  Observations
by ASCA (Tanaka 1993c; Moss 1997; Ebisawa 1997) and 
Mir-Kvant (Borozdin, Arefiev, \& Sunyaev 1993)
showed a hard power law spectrum, with no evidence for an ultrasoft excess.
Rapid intensity variations were detected by BATSE (Harmon et al. 1993)
and ASCA.  A strong QPO peak was apparent, with a centroid frequency that
varied from $\sim$0.04 Hz to $\sim$0.3 Hz through the discovery outburst
(van der Hooft et al. 1996).  A radio
(Mirabel, Rodriguez, \& Cordier 1993) 
and optical (Della Valle, Mirabel, \& Cordier 1993) 
low-mass counterpart has been identified at an estimated distance of 
$\simeq$2.4 kpc (Della Valle, Mirabel, \& Rodriguez 1994).  Interpreting
optical modulations at 14.7 h as a superhump period, Masetti et al. (1996)
estimated the mass of the compact object to be $>$4.9$\Msun$.  Hjellming
et al. (1996) reported a rapidly decaying radio flare following the
final outburst shown in Fig. \ref{lightcurves}.  This temporal coincidence
suggests a connection between the hard X-ray decline and the ejection
of the relativistic electrons responsible for the radio emission, and
from the similarity of this event to the emergence of a relativistic
radio-synchrotron jet in GRO~J1655--40 (Hjellming \& Rupen 1995),
Hjellming et al. suggested that a relativistic jet might have been present
in GRS~1716--249 at this time.  The radio data were inadequate to confirm
or reject this suggestion, unfortunately.

\begin{figure}
\vbox{\psfig{figure=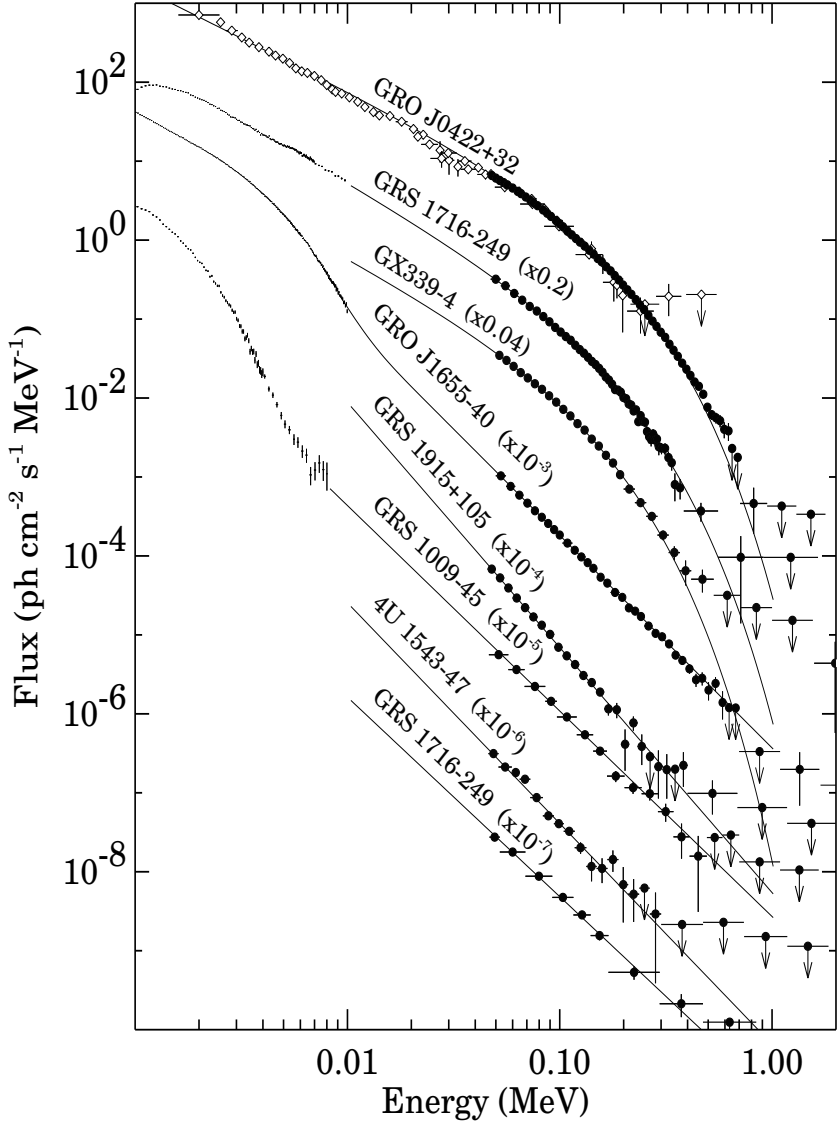,height=16.0cm,width=12.2cm}}
\caption{Photon number spectra from OSSE for seven transient
BHCs.  Spectra are averaged over all observing days for which there
was detectable emission
and, for clarity of the figure, have been scaled by arbitrary factors.
Two spectral states are apparent.
Contemporaneous TTM and HEXE data (open diamonds) and ASCA data (crosses)
are shown for GRO~J0422+32 and GRS~1716--249, respectively.
Non-contemporaneous ASCA data (crosses)
are shown for GRS~1009--45 and GRO~J1655--40, scaled as indicated in the text.}
\label{spectra}
\end{figure}

OSSE observed GRS 1716--249 five times, with the first two observations
beginning about 30 days after the onset of the discovery outburst, while the
20--100 keV flux was still within 10--15\% of its maximum intensity (these
observations are closely spaced in time and indistinguishable in
Fig. \ref{lightcurves}).  The remaining three observations were made on the
trailing edge of the second outburst, between the second and third
outbursts, and finally near the peak of the third outburst.  The gamma-ray
luminosity varied by an order of magnitude between observations.  There
is good evidence that both 
spectral states were observed.  The first, second, and fifth observations,
all made near the peak of outbursts, show the exponentially truncated
power law form of Eqn. 1, while the third is a simple power
law spectrum.  The spectrum of the fourth observation can be fit with
either functional form, and a statistically significant distinction cannot
be drawn.  The exponentially truncated spectra have been summed 
and are shown in the upper GRS~1716--249 spectrum in Fig. \ref{spectra},
while the third observation is shown in the lower spectrum.  
The spectral parameters in Table \ref{fits} are similarly
divided.  OSSE has detected spectral shape changes only in this BHC
and Cyg X-1 (Phlips et al. 1996).  Also plotted in Fig. \ref{spectra} is 
the X-ray spectrum from an ASCA observation on 1993 October 5 (Moss 1997), 
approximately three weeks before
the first OSSE observation and during the discovery outburst.  No ultrasoft
excess is apparent, which indicates that the source is in the X-ray low, hard
state.

\subsection{GRS~1915+105}

\begin{figure}
\vbox{\psfig{file=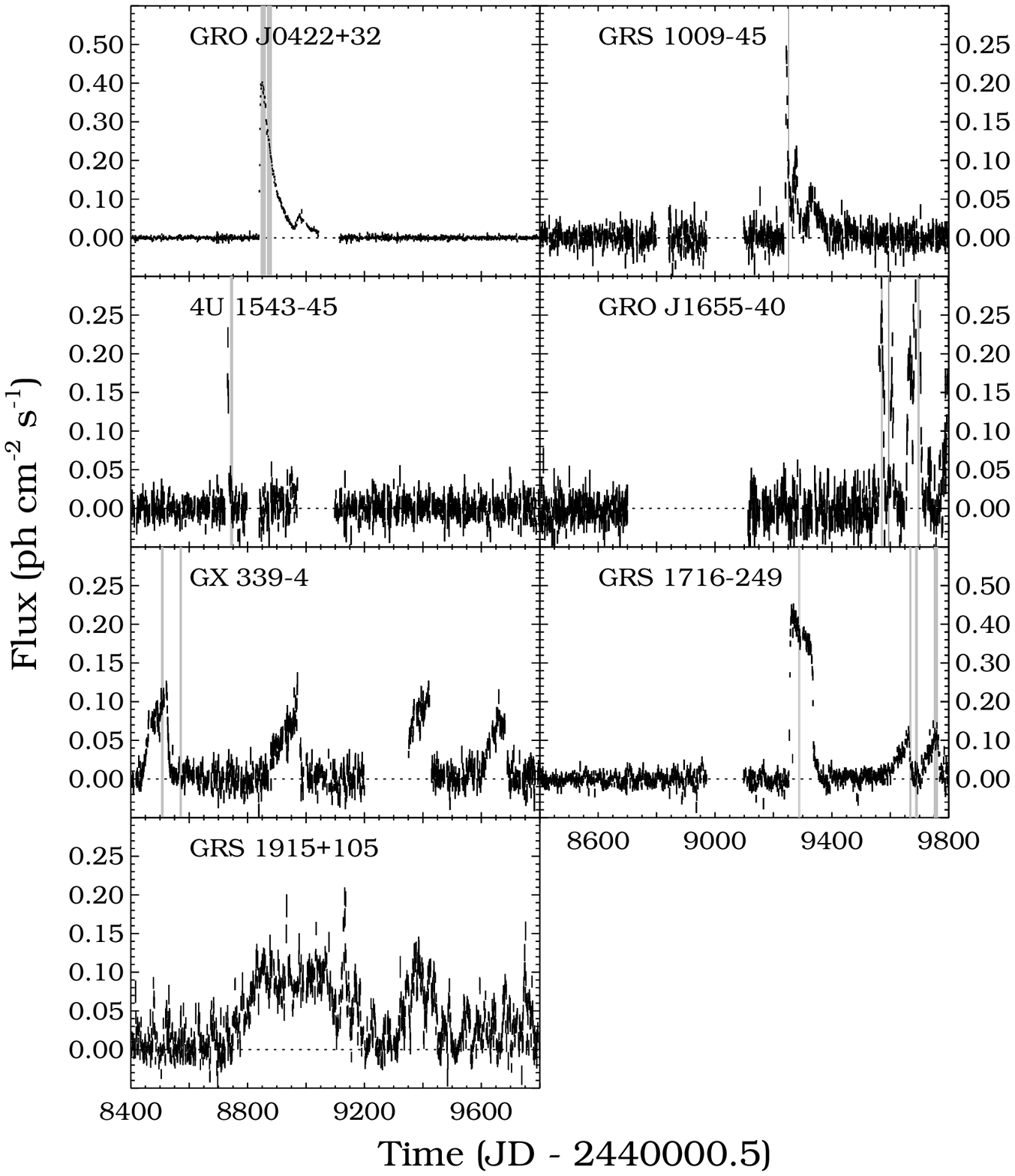,height=17.0cm,width=13.6cm}}
\caption{Lightcurves for seven transient BHCs 
in the 20--100 keV band as measured by BATSE.  Each data
point corresponds to a single day (JD 2448400.5 is 1991 May 24.0).  
The total Crab emission in this band is $\simeq$0.32 ph cm$^{-2}$
s$^{-1}$.  Days on which OSSE observed
each source are indicated by shaded regions.  Note that some OSSE observations
occurred following the time period covered in each panel (e.g. all
observations of GRS~1915+105).}
\label{lightcurves}
\end{figure}

The transient GRS 1915+105 (XN Aql 1992) was discovered by GRANAT/WATCH
on 1992 August 15 (Castro-Tirado et al. 1992).  The discovery outburst
began with a slow rise to maximum intensity, and since that time
its hard X-ray lightcurve has been episodic (Fig. \ref{lightcurves}).  On
timescales of tens to thousands of seconds, the X-ray emission is
dramatically variable, showing several repeating temporal structures
(Greiner, Morgan, \& Remillard 1996).  The X-ray spectrum is also
variable:  during intense flares the spectrum is cut off near 5 keV,
while during more stable periods it shows an ultrasoft disk blackbody
plus a power law with $\Gamma = -2.2$ (Greiner et al. 1996).
At a distance of $\simeq$12.5 kpc, it was 
the first object in our galaxy shown to have superluminal radio jets
(Mirabel \& Rodriguez 1994).  As with GRO~J1655--40, radio flares
frequently, although not always, follow hard X-ray flares and
occur during a decrease in the hard X-ray emission (Foster et al. 1996).
Mirabel et al. (1997) have suggested that the companion
is a high-mass late Oe or early Be star.  A mass function for the
system is not yet available.  

OSSE observed GRS~1915+105 on three occasions, all of which are
after the final day shown in Fig. \ref{lightcurves}.  Extrapolating the
best-fit power law model parameters from each of these viewing periods
into the 20--100 keV band gives 0.09, 0.07, and 0.07 ph cm$^{-2}$ s$^{-1}$, 
respectively, indicating that both observations were near the
historical peak hard X-ray intensity 
for this source (compare Fig. \ref{lightcurves}).
In all cases, the spectrum is a simple power law with no evidence for
a break.  As with GRO~J1655--40, there is a weak anti-correlation of
spectral hardness with $L_{\gamma}$ (i.e. as $L_{\gamma}$ increases,
the spectrum becomes softer and the photon number index becomes more
negative).  GRS~1915+105 was observed in 
1996 October by ASCA, Rossi XTE, Beppo SAX,
and OSSE in a scheduled campaign during a continuing outburst, so
simultaneous, high-quality, broadband spectra are
forthcoming for this object.

\subsection{Spectral Analysis}

Photon number spectra from each BHC
are shown in Fig. \ref{spectra}, along with the best-fit analytic model
extrapolated to 10 keV.  In most cases, the spectra are averaged over the
entire OSSE dataset for which the source was detected.  GRS~1716--249 appears
in this figure
twice, with the high-luminosity breaking 
spectra summed together, and the
low-luminosity power-law spectra summed together, as discussed 
below.  For clarity of the figure, the spectra 
have been multiplied by the arbitrary scaling factors indicated next to
the source name.  X-ray data are shown
for GRO~J0422+32, GRS~1009--45, GRO~J1655--40, and GRS~1716--249.

We fit the average spectra using one of two general analytic spectral models,
either a simple power law, or a power law that is
exponentially truncated above a break energy:
\begin{equation}
f(E) = \left\{ \begin{array}{ll}
    A \, E^{-\Gamma}  				&  \mbox{$E < E_b$} \\
    A \, E^{-\Gamma} \, \exp(-(E-E_b)/E_f) 	&  \mbox{$E > E_b$}
		  \end{array}
	 \right.
\end{equation}
where $A$ is the photon number flux, $\Gamma$ is the photon number
index, $E_b$ is
the break energy, and $E_f$ is the exponential folding energy.
Below the break energy, the exponential factor is replaced by unity, and
the model simplifies to a power law.  Best-fit model parameters
are given in Table \ref{fits}, along with the corresponding luminosities
in the gamma-ray band (i.e. above 50 keV).
Uncertainties in the model parameters are statistical only
and reported as 68\%-confidence intervals or 95\%-confidence lower limits.
If the simple power law is a statistically adequate fit, we report in
Table \ref{fits} the photon number index from that model along with
lower limits to $E_b$ and $E_f$.  Because of the strong correlation between
these two parameters when neither is required by the data, to establish
the lower limits we fixed $E_f = 2 E_b$, since that relation roughly holds
for GRO~J0422+32.

For four sources --- GRS~1009--45, 4U~1543--47, GX339--4, and GRS~1915+105
--- little change in spectral parameters is observed from day to day
or from observation to observation when the source is detected.
Larger changes are observed for GRO~J0422+32 
(where $E_f$ varies), GRO~J1655--40 (where $\Gamma$ varies), 
and GRS~1716--249 (where the spectral shape changes).
In Table \ref{fits} we also report the range of statistically significant
variability from day to day in one spectral parameter, $E_f$ for the
breaking spectra and $\Gamma$ for the power-law spectra,
along with the range of luminosities above 50 keV observed from day to day.

Two general spectral states in the gamma-ray band are apparent in
Fig. \ref{spectra}.  Three transients (GRO~J0422+32, GX339--4, and 
GRS~1716--249)
have breaking spectra.  The X-ray photon number
index is $\sim$1.5, and the exponential folding energy is
$\sim$100 keV.  The remaining four transients
(GRS~1009--45, 4U~1543--47, GRO~J1655--40, and GRS~1915+105)
show pure power law spectra, with a softer
photon number index, in the range $\simeq$2.5--3.  At low luminosity, 
GRS~1716-249 also appears to show a pure power law spectrum, with 
$\Gamma \simeq 2.5$.

The two spectral states have their peak luminosities in distinct energy
ranges.  Fig. \ref{nufnu} shows combined
X-ray and OSSE gamma-ray spectra for several sources, plotted as 
luminosity per logarithmic energy interval.  The sources shown in the 
breaking gamma-ray state, GRO~J0422+32 and GRS~1716--249, have the
bulk of their
luminosity emitted near 100 keV, while for GRO~J1655--40 in the
power-law gamma-ray state, the ultrasoft component is dominant.  We also
note that the distinction between the spectral states is more apparent
plotted in this manner.

We have searched for evidence of line emission at the energies
suggested in previous observations of
BHCs, i.e. at 170 keV from backscattered 511 keV
radiation, near 480 keV as reported from SIGMA data
on Nova Muscae (Goldwurm et al. 1992,
Sunyaev et al. 1992), and at 511 keV.
With OSSE's very high sensitivity (roughly
one order of magnitude greater than SIGMA's at 511 keV), previously
reported lines would be detected at high significance.
For example, a broadened 480 keV line at $6 \times 10^{-3}$ ph cm$^{-2}$
s$^{-1}$, the intensity reported from Nova Muscae,
would have been detected by OSSE at $\sim$40$\sigma$ in an average
24-hour period.  As the SIGMA data suggest, these lines might indeed
be variable, so we have searched our data on several timescales.
Table \ref{lines} 
summarizes the results of our search, reporting 
5$\sigma$ upper limits daily, weekly, and on
the total of all observations.  (We chose this high statistical
confidence level because of the large number of intervals searched.)
The observations in Table \ref{dates}
were divided into weeks of 7 contiguous days if possible; otherwise, if
the number of contiguous observing days was less than 7,
those contiguous days were defined as one ``week.''
No statistically significant narrow or moderately broadened line emission is 
observed at any time
from any of these objects.  

\section{Discussion}

\subsection{Is the Gamma-Ray State Correlated with Any Other
Observable?}

Having identified two distinct spectral states of gamma-ray emission
from galactic black hole transients, we are led to consider whether
the gamma-ray state is correlated with any other property
of the transient, such as the X-ray spectral state, the gamma-ray luminosity, 
the lightcurve of the outburst, 
the presence or absence of rapid variability, or the emission
in the radio band.  These properties are summarized in Table \ref{correlations}
for each of the BHCs.

There is a strong correlation between the gamma-ray spectra shape and the
existence of a strong ultrasoft X-ray spectral component.  In 
no case is a strong ultrasoft excess reported
during an outburst for which we find a breaking spectrum
in the gamma-ray band, and in all cases for which the gamma-ray
spectrum is a power law, a strong ultrasoft excess {\em is} reported in
the X-ray band at some time during the outburst (other than for GRS~1716--249,
for which there appears to be no contemporaneous X-ray data).  We 
are lead to identify the power law gamma-ray state with the X-ray
high, soft state (or perhaps the X-ray very high state) and the
breaking gamma-ray state with the X-ray low, hard state.
In the former case, the bulk of
the luminosity is carried off in $\sim$1 keV photons from the ultrasoft
component, while in the latter case, it is carried by $\sim$100 keV photons
(see Fig. \ref{nufnu}).  
We note that Ebisawa, Titarchuk, \& Chakrabarti (1996) reported
from Ginga data that the 
photon number index of the power-law tail is always greater than 2 in
the X-ray high, soft state and always less than 2 in the X-ray low, hard
state.  Our OSSE observations are consistent with this behavior; thus
we are led to suggest that a measurement of the X-ray spectral index 
is adequate to indicate the presence or absence of the ultrasoft X-ray
component.  We caution, however, that
the number of truly simultaneous X-ray and gamma-ray observations
of BHCs is small, and further broadband observations, 
particularly with ASCA, Rossi
XTE, or Beppo SAX and with OSSE, are required to study the 
correlation between X-ray and gamma-ray states in detail.

There is no apparent correlation between gamma-ray luminosity (above 50 keV)
and the spectral state from source to source, since both the highest and
lowest luminosity objects, GRS~1915+105 and 4U~1543--47 respectively,
have power-law gamma-ray spectra.  However, in the case of the one source
that appears to show a state change, GRS~1716--249, the power-law state
occurs when $L_{\gamma}$ is low.  This is also the case for Cyg X--1
(Phlips et al. 1996).  Thus it appears to be the case that, for an
individual BHC, the power-law spectral state has lower
$L_{\gamma}$ than does the breaking state.  How the
{\em bolometric} luminosity changes is less certain and will require
further simultaneous broadband observations.

Harmon et al. (1994) identified two types of black hole transients based on
their lightcurves above 20 keV.  The first type has a relatively fast rise
followed by a more gradual decay, frequently with secondary maxima
some weeks or months into the decline.  The decay sometimes 
approaches an exponential form as seen in the soft X-ray 
band (Tanaka and Lewin 1995).  The rise and decay times vary over
a broad range, but are of the order of a few days and a few tens of days,
respectively.  We have labeled transients of this type as ``FRED'', for
fast rise and exponential decay,
in Table \ref{correlations}.  Harmon et al. use the label ``X-ray novae.''
The second type has a longer rise time
(of order weeks) and multiple, recurrent outbursts of highly variable
duration.  Major outbursts tend to be of similar intensity, with 
active periods lasting for several years.   We have labeled this type
as ``episodic'' in Table \ref{correlations}.  Harmon et al. call them the
``slow risers.''  No coherent, long-term periodicities in the
high energy emission have been
seen in either type of transient.  We find that there is no
correlation between the lightcurve type and the gamma-ray spectral state.
The transients with FRED lightcurves
clearly show both spectral states (e.g. compare
GRO~J0422+32 and GRO~J1009--45), and the transients in the breaking
state, GRO~J0422+32
and GRS~1716--249, have clearly distinct lightcurves.  Furthermore, Cyg X-1
and GRS~1716--249 exhibit {\em both} gamma-ray spectral
states.  Thus the mechanism
that gates the accretion flow, and therefore regulates the production 
and evolution of
outbursts, does not determine the physical process responsible for the
gamma-ray emission.

Strong, rapid, aperiodic 
variability (i.e. on timescales of tens of seconds or less)
is frequently reported in X-rays from BHCs in the X-ray low, hard state.
Fractional rms variability is typically of order tens
of percent of the average intensity.  For recent 
reviews, see van der Klis (1994, 1995).
We find that this is generally true in the gamma-ray band as well; see
e.g. Grove et al. (1994) for GRO~J0422+32 and van der Hooft et al.
(1996) for GRS~1716--249, where in both sources
strong, rapid variability and peaked noise are detected in the
breaking gamma-ray state.  The strong variability in both
energy bands confirms the correspondence
between the breaking gamma-ray state and the X-ray low, hard state.
Recent results from RXTE 
(e.g. Morgan, Remillard, \& Greiner 1997) indicate that the 
superluminal sources GRO~J1655--40 and
GRS~1915+105 in the X-ray high, soft state 
both show weaker rapid X-ray variability (of order several percent rms). 
While no such
variability is detected in the gamma-ray band, the statistical
upper limits are $\sim$5\% and thus reasonably consistent with the X-ray result 
(Crary et al. 1996, Kroeger et al. 1996).

For the superluminal sources GRO~J1655--40 and GRS~1915+105, correlation 
between radio flaring and decreases in gamma-ray intensity has been noted 
elsewhere (Foster et al. 1996, Harmon et al. 1997).  Yet this correlation
is not entirely consistent:  some radio flares are not followed by gamma-ray
decreases, and perhaps more importantly, while the discovery outburst of
GRO~J1655--40 was characterized by multiple gamma-ray dips and flares
accompanied by radio flares and knot ejections, later gamma-ray outbursts
showed little or no radio variability (Tavani et al. 1996).  We find
that the gamma-ray
spectral state of the superluminal sources remains unchanged within an
outburst and from one outburst to the next, despite the presence
or absence of a subsequent radio flare or ejection episode.  Thus,
while there certainly is a correlation between gamma-ray
and radio events, the nature and details of the physical connection
remain obscure.

\subsection{Comparison with Cyg X-1}

The archetypal binary for characterization and modeling of high-energy
emission from black hole systems is Cyg X-1.  The compact object has
a mass in excess of 6$\Msun$ (Gies \& Bolton 1986), and the companion
is a massive O-type supergiant in a 5.6-day binary orbit.
Cyg X-1 is commonly in the X-ray low, hard state, and on occasion transitions
to the X-ray high, soft state.  Ling et al. (1987) enumerated three
distinct levels in the 45--140 keV gamma-ray flux during the X-ray
low, hard state, while Bassani et al. (1989) noted that the intensity
of the hard X-ray flux might be anti-correlated with the soft X-ray
flux.  In the X-ray low, hard state, the source is
radio-bright and relatively stable, while in the X-ray high, soft state,
it is radio-quiet (Tananbaum et al. 1972, Hjellming 1973).  The persistence
of the radio emission along with weak modulation apparently at the binary
period suggest the presence of an optically thick jet (Hjellming \& Han 1995).

Recent broadband observations of Cyg X-1 reveal
a bimodal spectral behavior in the gamma-ray
band as well as the X-ray band (Phlips et al. 1996).  At high $L_\gamma$,
the spectrum is exponentially breaking with
folding energy $\simeq$100 keV, while at low $L_\gamma$, 
the spectrum is characterized
by a power law with $\Gamma \simeq 2.3$.  Thus Cyg X-1 fits neatly in
the categories developed here for transient black holes.  Furthermore,
recent contemporaneous measurements with OSSE and Ginga or RXTE in both
X-ray and gamma-ray states reveal that indeed the X-ray high, soft state
is identified with the power law gamma-ray state, and the X-ray
low, hard state is identified with the breaking gamma-ray state,
a behavior equivalent to that shown in Fig. \ref{nufnu}
(Gierlinski et al. 1997; Cui et al. 1997; Phlips et al. 1998, in preparation).

\subsection{Emission Mechanisms}

In recent years there has been considerable interest in advection-dominated
accretion disk models (Narayan \& Yi 1994; 1995a; 1995b; Abramowicz et al.
1995; Chen et al. 1995). In these models radiative cooling is inefficient, and
the bulk of the gravitational binding energy liberated through viscous
dissipation is advected across the event horizon of the black hole. Stable
global solutions exist  (Narayan, Kato, \& Honma 1997; Chen, Abramowicz, \&
Lasota 1997) in which advective cooling dominates in the inner region where a
hot optically thin two-temperature plasma is formed. This joins to a standard
thin disk solution (Shakura \& Sunyaev 1973; see also Frank, King, \& Raine
1992) at large radii. The gamma ray emission is produced in the inner region
through thermal
Comptonization of soft photons, the bulk of which are probably supplied
by the cool outer disk.

Ebisawa, Titarchuk, \& Chakrabarti (1996) propose that the two distinct 
spectral indices in the hard X-ray band, corresponding to the X-ray
high, soft and X-ray low, hard states, are the result of two distinct
Comptonization mechanisms.  In the low, hard state, the hard X-ray and
gamma-ray emission is produced by thermal Comptonization of soft photons
from the accretion disk.  By contrast, in the high, soft state, the
high-energy emission arises from bulk-motion Comptonization 
in the convergent accretion flow from the inner edge of the accretion disk.
We find that the broadband X-ray and gamma-ray spectra in the X-ray
high, soft and low, hard states are consistent with this picture.
The spectra shown in Fig. \ref{spectra} suggest 
the general scenario in which the
presence of copious soft photons from the disk in the high, soft state serves
to cool the electrons efficiently in the inner advection-dominated
Comptonization region. In this case, Comptonization due to bulk motion dominates
over that due to thermal motion.  In contrast, in the low, hard state, there
are fewer soft photons, hence higher temperatures in the 
Comptonization region (i.e. $\sim$50--100 keV, rather than $\sim$1 keV), and
thermal Comptonization dominates.

We note that Ebisawa et al. (1996) predict
that the power law spectrum from bulk-motion Comptonization will be sharply
cut off near the electron rest mass-energy, $m_e c^2$,
somewhere in the 200--500 keV
band, because the efficiency of Comptonization drops due to increasing
electron recoil as the photon energy increases.
However, a recalculation by Titarchuk, Mastichiadis, \& Kylafis (1996),
including 
the term dependent on the second order in the velocity of the bulk flow,
shows that the cutoff energy is substantially increased, to
beyond $m_e c^2$, without affecting the power-law index.
Thus the extended power-law spectrum from GRO~J1655--40 does indeed appear
to be consistent with bulk-motion Comptonization.

\begin{figure}
\vbox{\psfig{figure=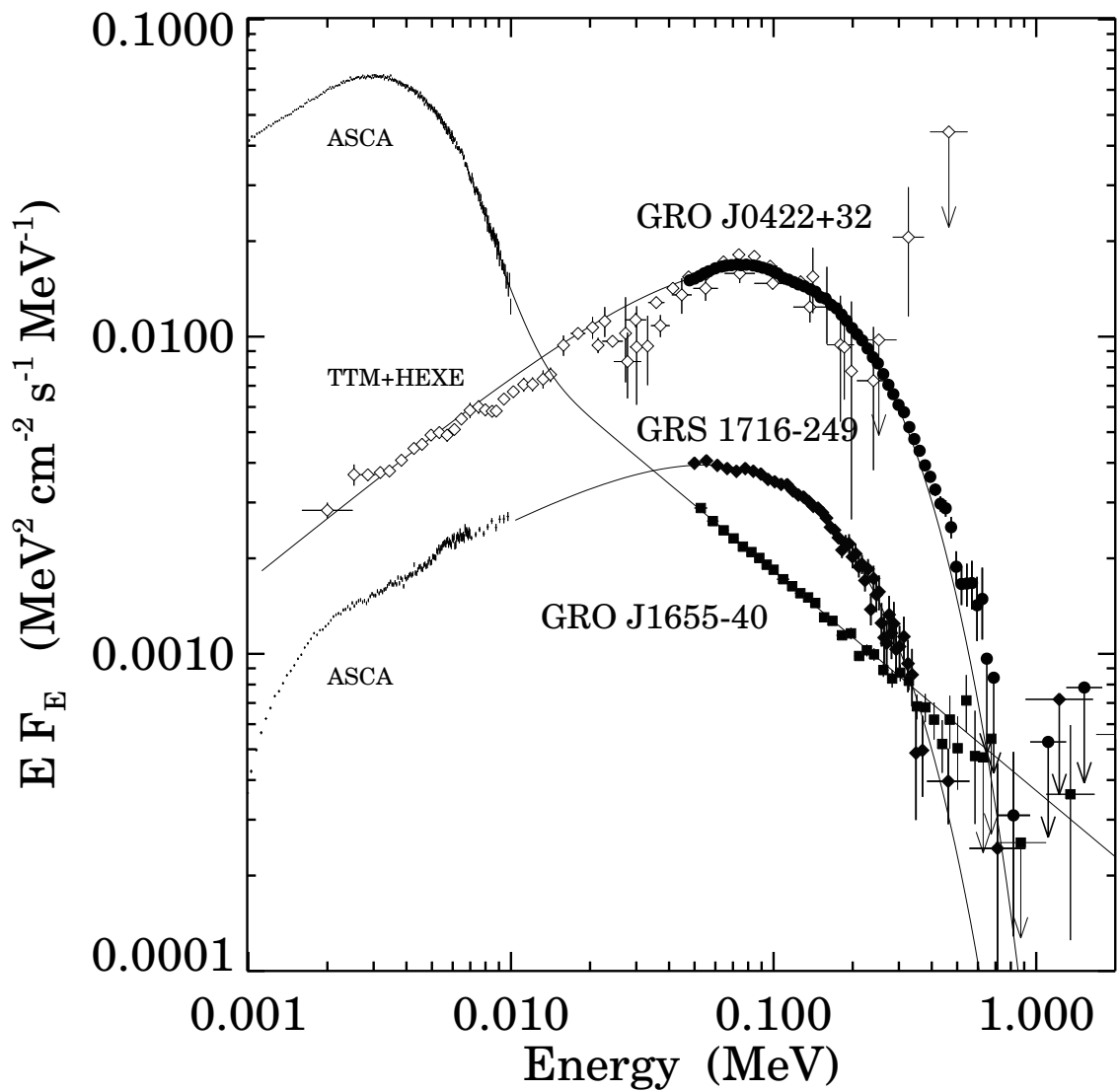,height=16.0cm,width=16.0cm}}
\caption{Luminosity per logarithmic energy interval for those transient
BHCs with best-quality broadband spectral data.
Contemporaneous TTM and HEXE data (open diamonds) and ASCA data (crosses)
are shown for GRO~J0422+32 and GRS~1716--249, respectively.
Non-contemporaneous ASCA data
are shown for GRO~J1655--40 (crosses), scaled as indicated in the text.}
\label{nufnu}
\end{figure}

Recently Narayan, Garcia, \& McClintock (1997) have compared the long-term
X-ray
luminosity variations of neutron star and black hole X-ray transients and have
found that the black hole transients display much larger luminosity variations
between quiescence and outburst than do neutron stars.  
They attributed this to the fact that black holes have horizons, whereas
neutron stars have material surfaces.  Their study is based upon the 
luminosities measured in the 0.5--10 keV band using sources in outburst
in both the
X-ray low, hard state (e.g. V404 Cyg) and the X-ray high, soft state
(e.g. A0620--00).  It is evident from Fig. \ref{nufnu} that in fact the
total luminosity in the X-ray low, hard state is substantially higher
than the 0.5--10 keV luminosity, and therefore that the true variation
between quiescence and outburst in these sources is likely to be
substantially greater than Narayan et al. estimate.  Indeed the spectra
in Fig. \ref{nufnu} indicate that the luminosity above 50 keV (10 keV) exceeds
that in the 0.5-10 keV band by a factor of $\simeq$4 ($\simeq$6).
Unfortunately, no gamma-ray detections of BHCs in the quiescent state
exist at this date, and therefore true measurements of the variation in
bolometric luminosity cannot yet be made.

The observations we report here of transients in the X-ray low, hard
state are entirely consistent with high-energy emission from thermal
Comptonization.  The spectra smoothly turn over 
with e-folding energy $E_f\sim100$ keV, which
implies an electron temperature of $kT\lesssim100$ keV for the Comptonizing
plasma. The spectral indices of the power law portion are $\Gamma\simeq1.5$,
which at temperatures $\sim100$ keV imply Thompson depths $\tau_T\gtrsim1$ for
scattering in a spherical plasma cloud (Skibo et al. 1995).

As is the case for Cyg X-1 (Gierlinski et al. 1997), the source GRO~J0422+32
appears to be ``photon starved'' in that the gamma ray luminosity 
greatly exceeds the
soft X-ray luminosity.  This starvation and, in the case of
GRO~J0422+32, the apparent absence of a reflection component 
argue against the disk-corona geometry (Haardt \& Maraschi 1993), but
are consistent with a geometry where the Comptonizing region forms a
quasi-spherical cloud near the black hole as predicted by advection-dominated
models.

\section{Conclusions}

We have demonstrated the existence of two distinct spectral states 
of gamma-ray emission from galactic BHCs.  The first state
exhibits a relatively hard power-law component with photon index roughly 1.5 
that breaks at $\simeq$50 keV and is exponentially truncated with an e-folding
energy of $\simeq$100 keV.  This ``breaking gamma-ray''
is firmly identified with the 
X-ray low, hard state, i.e. with the absence or weakness
of an ultrasoft X-ray component.
Sources exhibiting this spectral signature include
GRO~J0422+32 and GRS~1716--249, as well as GX~339--4 and Cyg X-1 in their
X-ray low, hard states.  The second state exhibits a relatively soft
power-law gamma-ray spectrum with photon index roughly 2--3 and 
no evidence of a break.  In the case of GRO~J1655--40, the lower limit
on the break energy is 690 keV, above the electron rest mass.
This ``power law gamma-ray'' state is identified with the presence
of an ultrasoft X-ray component, which is the signature of the X-ray
high, soft state and the X-ray very high state.  Because of the paucity
of simultaneous X-ray and gamma-ray observations and the difficulty of
distinguishing between the two X-ray high states, the identification
of the power law gamma-ray state with which of the two high states
is somewhat uncertain, although Cyg~X-1 observations indicate that the
correspondence is with the high, soft state (Cui et al. 1997).
Sources exhibiting the power law gamma-ray state include
GRS~1009-45, 4U~1543--47, GRO~J1655--40, and GRS~1915+105, as well as
Cyg~X-1 in its X-ray high, soft state.  Gamma-ray spectral state changes
have now been observed in GRS~1716--249 and Cyg~X-1.  Further 
simultaneous X-ray and gamma-ray spectroscopy measurements of galactic BHCs 
will help elucidate the details of the relationship of the emission in
the two bands.

\acknowledgements
This work was supported by NASA contract S-10987-C.

\clearpage

\begin{table}
\begin{center}
\begin{tabular}{ll}
\tableline
\tableline
Object		&	Observation Dates \\
\tableline
& \\
GRO~J0422+32	&	1992 Aug 11--27, 1992 Sep 1--17 \\
GRS~1009-45	&	1993 Sep 21--22 \\
4U~1543--47	&	1992 Apr 28 -- May 7 \\
GRO~J1655--40   &       1994 Aug 4--9, 1994 Dec 7--13, \\
                &       1995 Mar 29 -- Apr 4, 1996 Aug 27 -- Sep 6 \\
GX~339--4       &       1991 Sep 5--12 \\
GRS~1716--249   &       1993 Oct 25--27\tablenotemark{a}, 
    1993 Oct 30--31\tablenotemark{a}, \\
                &       1994 Nov 9--15\tablenotemark{b}, 
    1994 Nov 29 -- Dec 7\tablenotemark{c} \\
		&	1995 Feb 1--14\tablenotemark{a} \\
GRS~1915+105	&	1995 Nov 21 -- Dec 7, 1996 Oct 15 -- 29, \\
		&	1997 May 14--20 \\
\tableline
\end{tabular}
\end{center}
\caption{OSSE observation list}
\tablenotetext{a}{Breaking gamma-ray state.}
\tablenotetext{b}{Power-law gamma-ray state.}
\tablenotetext{c}{Spectral state uncertain; data not used.}
\label{dates}
\end{table}

\clearpage

\begin{table*}
\begin{center}
\scriptsize
\begin{tabular}{lccccccc}
\tableline
\tableline
Object & $< \Gamma >$ & $\Delta \Gamma$ & $<E_{break}>$ & $<E_{fold}>$ & 
  $\Delta E_{fold}$ & $<L_\gamma >$ & $\Delta L_\gamma$ \\
& & & (keV) & (keV) & (keV) & ($\times 10^{36}$ erg/s) & ($\times 10^{36}$ erg/s) \\
\tableline
& & & & & & & \\
\multicolumn{8}{l}{\it Breaking gamma-ray state} \\
GRO~J0422+32  & 1.49$\pm$0.01\tablenotemark{a} & & 60$\pm$3\tablenotemark{a} 
    & 132$\pm$2\tablenotemark{a} & 120--155 & 19. & 14.--26. \\
GX339--4      & 1.38$\pm$0.08 & & $<$50    &  87$\pm$6 & 80--90   & 5.4 &
8.7--9.3 \\
GRS~1716--249 & 1.53$\pm$0.06 & & $<$50    & 115$\pm$8 & 80--140  & 6.0 &
2.3--11.2 \\
\tableline
& & & & \\
\multicolumn{8}{l}{\it Power-law gamma-ray state} \\
GRS~1009-45   & 2.40$\pm$0.06 & $\simeq$2.4 & $>$160 & $>$320 & & 3.9 & 
$\simeq$3.9 \\
4U~1543--47   & 2.78$\pm$0.05 & $\simeq$2.8 & $>$200 & $>$400 & & 0.8 & 
0.5--1.7 \\
GRO~J1655--40 & 2.76$\pm$0.01 & 2.6--3.0    & $>$690 & $>$1380 & & 6.1 & 
5.1--9.5 \\
GRS~1716--249 & 2.42$\pm$0.08 & $\simeq$2.4 & $>$250 & $>$500 & & 1.1 & 
0.6--1.7 \\
GRS~1915+105  & 3.08$\pm$0.06 & 2.9--3.3    & $>$390 & $>$780 & & 31. & 
27.--45. \\
\tableline
\end{tabular}
\end{center}
\caption{Best-fit model parameters}
\tablenotetext{a}{OSSE data from 1992 Aug 26--27 and Sep 1--2 only.}
\tablecomments{Fits are
performed over sum of entire data set for each object for which there
was detectable emission, except for GRO~J0422+32, where the fit is for the
contemporaneous HEXE and OSSE observation, and for GRS~1716--249, where the
fit is performed separately for the sum of observations in each spectral
state (see text).  Errors are 68\% confidence.  Lower limits on 
exponential folding energies are
95\% confidence.  Upper limits on break energies are set at OSSE threshold
for spectroscopy.  Luminosity is for energies above 50 keV.  Ranges
for spectral index,
folding energy, and luminosity from daily fits are given in the columns
labeled by $\Delta \Gamma$, $\Delta E_{fold}$, and $\Delta L_\gamma$.}
\label{fits}
\end{table*}

\clearpage

\begin{table*}[tbh]
\begin{center}
\scriptsize
\begin{tabular} {lccccc}
\\
\tableline
\tableline
Source & Line & FWHM & \multicolumn{3}{c}{Flux ($\times 10^{-4}$ ph cm$^{-2}$ 
s$^{-1}$)}\tablenotemark{a} \\
 & Centroid (keV) & (keV) & Daily & Weekly\tablenotemark{b} & Total \\
\tableline
GRO J0422+32 & 170 &  3 & $<$6.3 & $<$2.3 & $<$1.1 \\
	     & 170 & 35 & $<$9.3 & $<$3.5 & $<$1.6 \\
	     & 480 & 25 & $<$7.5 & $<$2.8 & $<$1.2 \\
	     & 511 &  3 & $<$7.2 & $<$2.7 & $<$1.2 \\
\tableline
GRS 1009--45 & 170 &  3 & $<$6.3 & $<$4.5 & $<$4.5 \\
	     & 170 & 35 & $<$8.9 & $<$6.5 & $<$6.5 \\
	     & 480 & 25 & $<$7.8 & $<$5.8 & $<$5.8 \\
	     & 511 &  3 & $<$7.7 & $<$5.7 & $<$5.7 \\
\tableline
4U 1543--47 & 170 &  3 & $<$5.4 & $<$1.7 & $<$1.7 \\
	    & 170 & 35 & $<$7.6 & $<$2.4 & $<$2.4 \\
	    & 480 & 25 & $<$7.1 & $<$2.3 & $<$2.3 \\
	    & 511 &  3 & $<$6.8 & $<$2.3 & $<$2.3 \\
\tableline
GRO J1655--40 & 170 &  3 & $<$5.1 & $<$2.2 & $<$0.9 \\
	      & 170 & 35 & $<$7.2 & $<$3.1 & $<$1.3 \\
	      & 480 & 25 & $<$6.2 & $<$2.6 & $<$1.1 \\
	      & 511 &  3 & $<$6.1 & $<$2.6 & $<$1.0 \\
\tableline
GX 339--4 & 170 &  3 & $<$5.7 & $<$2.0 & $<$1.3 \\
	  & 170 & 35 & $<$8.7 & $<$3.1 & $<$1.9 \\
	  & 480 & 25 & $<$6.1 & $<$2.3 & $<$1.4 \\
	  & 511 &  3 & $<$6.1 & $<$2.2 & $<$1.3 \\
\tableline
GRS 1716--249 & 170 &  3 & $<$10.2 & $<$4.0 & $<$2.1 \\
	      & 170 & 35 & $<$15.0 & $<$5.9 & $<$3.1 \\
	      & 480 & 25 & $<$11.5 & $<$5.0 & $<$2.6 \\
	      & 511 &  3 & $<$10.3 & $<$4.6 & $<$2.4 \\
\tableline
GRS 1915+105 & 170 &  3 & $<$5.8 & $<$2.0 & $<$1.2 \\
	     & 170 & 35 & $<$7.9 & $<$2.8 & $<$1.6 \\
	     & 480 & 25 & $<$6.9 & $<$2.4 & $<$1.4 \\
	     & 511 &  3 & $<$6.8 & $<$2.4 & $<$1.8 \\
\tableline
\\
\end{tabular}
\end{center}
\caption{Upper limits on narrow and broad line emission}
\tablenotetext{a}{Upper limits are 5$\sigma$ ($\Delta\chi^{2} = 25$).}
\tablenotetext{b}{Denotes an interval of one to ten
contiguous days, depending on the source (see text).}
\label{lines}
\end{table*}

\clearpage

\begin{table*}
\begin{center}
\scriptsize
\begin{tabular}{lcccccc}
\tableline
\tableline
Object	      & Ultrasoft  & $L_\gamma $ ($>$50 keV)  & Lightcurve	
& \multicolumn{2}{c}{Timing Noise?} & Radio \\
	      & Component? & ($\times 10^{36}$ erg/s) & Type	
& $<$20 keV & $>$50 keV    & Emission? \\
\tableline
& & & & & & \\
\multicolumn{7}{l}{\it Breaking gamma-ray state} \\
GRO~J0422+32  & No\tablenotemark{a}  & 14.--26.  & FRED     & ? & Strong & 
  Transient\tablenotemark{b} \\
GX339--4      & No\tablenotemark{c}  & 8.7--9.3  & Episodic & 
  Strong\tablenotemark{c} & Weak & Variable; jet?\tablenotemark{d} \\
GRS~1716--249 & No\tablenotemark{e}  & 2.3--11.2 & Episodic & 
  Strong\tablenotemark{f} & Strong & Transient; jet?\tablenotemark{g} \\
\tableline
& & & & & & \\
\multicolumn{7}{l}{\it Power-law gamma-ray state} \\
GRS~1009--45  & Yes\tablenotemark{h} & $\simeq$3.9 & FRED     & 
  Weak\tablenotemark{f} & UL & ? \\
4U~1543--47   & Yes\tablenotemark{i} & 0.5--1.7    & FRED     & ?    & UL & ? \\
GRO~J1655--40 & Yes\tablenotemark{j} & 5.1--9.5    & Episodic & 
  Moderate\tablenotemark{k} & UL & SL jet\tablenotemark{l} \\
GRS~1716--249 & ?   & 0.6--1.7    & Episodic & ?    & UL & 
  Jet?\tablenotemark{g} \\
GRS~1915+105  & Yes\tablenotemark{m} & 27.--45.    & Episodic & 
  Moderate\tablenotemark{k} & UL & SL jet\tablenotemark{n} \\
\tableline
\end{tabular}
\end{center}
\caption{Phenomenological summary of black-hole transients.}
\tablerefs{(a) Tanaka 1993a; (b) Shrader et al. 1994; (c) Miyamoto et al. 1991;
(d) Sood \& Campbell-Wilson 1994; Fender et al. 1997; (e) Tanaka 1993c;
Borozdin, Arefiev, \& Sunyaev 1993; (f) Ebisawa 1997; (g) Hjellming \& Rupen
1995; Hjellming et al. 1996; (h) Kaniorsky, Borozdin, \& Sunyaev 1993;
Tanaka 1993b; (i) Matilsky et al. 1972; Greiner et al. 1993; (j) Zhang et
al. 1997; (k) Morgan, Remillard, \& Greiner 1997; (l) Tingay et al. 1995;
Hjellming \& Rupen 1995; (m) Greiner et al. 1996; (n) Mirabel \& Rodriguez
1994.}
\tablecomments{  ``FRED'' denotes a fast rise and 
exponential decay lightcurve.  The symbol ``?'' indicates
no reports in the literature.  ``UL'' indicates no evidence
at the sensitivity of the OSSE observations; only an upper limit is
available.  ``SL jet'' indicates apparent superluminal radio jet.}
\label{correlations}
\end{table*}

\clearpage

\clearpage

\clearpage

\end{document}